\documentclass[useAMS,usenatbib]{mn2e}
\usepackage{aas_macros,graphicx,times,multirow,amsmath}
\title[The X-ray emission of \src]{The X-ray emission of the high-mass X-ray binary \src}
\author[P.~Esposito et al.] {P.~Esposito,$^{1}$\thanks{E-mail: paoloesp@iasf-milano.inaf.it} G.~L.~Israel,$^{2}$ L.~Sidoli,$^1$ A.~Tiengo,$^{1,3,4}$ S.~Campana$^{5}$ and A.~Moretti,$^{6}$ 
\smallskip\\
$^1$Istituto di Astrofisica Spaziale e Fisica Cosmica - Milano, INAF, via E. Bassini 15, I-20133 Milano, Italy\\
$^2$Osservatorio Astronomico di Roma, INAF, via Frascati 33, I-00040 Monteporzio Catone, Italy\\
$^3$IUSS -- Istituto Universitario di Studi Superiori, piazza della Vittoria 15, I-27100 Pavia, Italy\\
$^4$INFN -- Istituto Nazionale di Fisica Nucleare, Sezione di Pavia, via A. Bassi 6, I-27100 Pavia, Italy\\
$^5$Osservatorio Astronomico di Brera, INAF, via E. Bianchi 46, I-23807 Merate, Italy\\
$^6$Osservatorio Astronomico di Brera, INAF, via Brera 28, I-20121 Milano, Italy
}
\date{Accepted 2014 April 2.  Received 2014 March 14; in original form 2014 February 16} \pagerange{\pageref{firstpage}--\pageref{lastpage}} \pubyear{2014}

\def\LaTeX{L\kern-.36em\raise.3ex\hbox{a}\kern-.15em
    T\kern-.1667em\lower.7ex\hbox{E}\kern-.125emX}
\def\xmm {\emph{XMM--Newton}}
\def\cxo {\emph{Chandra}}
\def\swift {\emph{Swift}}

\def\igr {\emph{INTEGRAL}}

\def\xte {\emph{RXTE}}
\def\rst {\emph{ROSAT}}

\def\src {IGR\,J17200--3116}
\def\flux {\mbox{erg cm$^{-2}$ s$^{-1}$}}
\def\lum {\mbox{erg s$^{-1}$}}
\def\nh {$N_{\rm H}$}

\begin{document}

\label{firstpage}
\maketitle
\begin{abstract}
The source \src\ was discovered in the hard X-ray band by \igr. A periodic X-ray modulation at $\sim$326~s was detected in its \swift\ light curves by our group (and subsequently confirmed by a \swift\ campaign). In this paper, we report on the analysis of all the \swift\ observations, which were collected between 2005 and 2011, and of an $\sim$20~ks \xmm\ pointing that was carried out in 2013 September. During the years covered by the \swift\ and \xmm\ observations, the 1--10~keV fluxes range from $\sim$1.5 to $4\times$10$^{-11}$~\flux. \src\ displays spectral variability as a function of the pulse phase and its light curves show at least one short (a few hundreds of seconds) dip, during which the flux dropped at 20--30\% of the average level. Overall, the timing and spectral characteristics of \src\ point to an accreting neutron star in a high-mass system but, while the pulse-phase spectral variability can be accounted for by assuming a variable local absorbing column density, the origin of the dip is unclear. We discuss different possible explanations for this feature, favouring a transition to an ineffective accretion regime, instead of an enhanced absorption along the line of sight.
\end{abstract}

\section{Introduction}

The source \src\ \citep{revnivtsev04short,walter04short} was discovered in 2003 during a survey of the Galactic Centre region with \igr. The hard X-ray (18--60~keV) flux of \src\ was $\sim$1.6~mCrab, roughly corresponding to $1.4\times10^{-11}$~\flux. \citet{revnivtsev04short} and \citet{stephen05short} pointed out that this new \igr\ source was the counterpart of the soft X-ray \rst\ point source 1RXS\,J172006.1--311702. \citet{masetti06short} proposed an optical counterpart 
with a narrow H$\alpha$ line and reddened continuum, suggesting a high-mass X-ray binary (HMXB) system. The association was fortified by \citet{tomsick08}, who obtained a refined \cxo\ X-ray position.

More recently, our team reported on the discovery of a coherent modulation of the X-ray emission of \src\ at a period of $\sim$326~s \citep{nichelli11}. The signal was detected in a run of the \swift\ Automatic Timing ANAlysis of Serendipitous Sources at Brera And Roma astronomical observatories (SATANASS\,@\,BAR) project. SATANASS\,@\,BAR consists in a systematic Fourier-based search for new pulsators in the \swift\ X-ray data.\footnote{For our analogous \cxo\ project, the \cxo\ ACIS Timing Survey at Brera And Rome astronomical observatories (CATS\,@\,BAR), see  \citet{eis13,eisrc13,eism13}.} So far, about 4000 X-ray Telescope (XRT) light curves of point sources with a sufficiently high number of photons ($\ga$150) were analysed and the effort yielded six previously unknown X-ray pulsators (including \src, two other new pulsators were reported in \citealt{nichelli09}). 

The original detection of the period of \src\ occurred in a couple of \swift\ consecutive observations performed in 2005 October 26--27. After that, more \swift\ observations were carried out between 2010 October and 2011 May to characterize better \src. The period measured during the new \swift\ campaign was slightly longer ($\sim$328~s; \citealt{nichelli11}, see also Section\,\ref{timingan}), indicating that the modulation reflects almost certainly the spin of a neutron star.
Here we give more details on the \swift\ observations and we report on a 20-ks long \xmm\ observation of \src\ performed on 2013 September 19, where we discovered an evident dip (a flux drop). While in low mass X-ray binaries these features are usually due to obscuring matter located in the outer accretion disc (see e.g. \citealt{diaztrigo06} for a review), in HMXBs they are probably produced by a transition to a different accretion regime (see \citealt{drave13} and references therein). In Sections \ref{observations} and \ref{analysis}, we describe the X-ray observations used and present the results of our timing and spectral analysis. Discussion follows in Section\,\ref{discussion}, where we concentrate on the origin of a peculiar short off-state observed in \src\ with \xmm.

\section{Observations}\label{observations}
\begin{table*}
\begin{minipage}{10.2cm}
\centering \caption{Summary of the \swift\ and \xmm\ observations used in this work.} \label{logs}
\begin{tabular}{@{}lcccc}
\hline
Mission~/~Obs.\,ID & Instrument & Date & Exposure & Net counts$^{a}$ \\
 & & & (ks) & \\
\hline
\swift~/~00035088001 & XRT & 2005 Oct 26 & 6.3 & $2312\pm48$\\
\swift~/~00035088002 & XRT & 2005 Oct 27 & 5.1 & $1350\pm37$ \\
\swift~/~00035088003 & XRT & 2010 Oct 22--23 & 10.5 & $2910\pm54$ \\
\swift~/~00035088004 & XRT & 2010 Oct 24 & 4.6 & $1105\pm33$ \\
\swift~/~00035088005 & XRT & 2010 Oct 27 & 5.3 & $886\pm30$ \\
\swift~/~00035088006 & XRT & 2011 Feb 03 & 10.1 & $2108\pm46$\\
\swift~/~00035088007 & XRT & 2011 Feb 05 & 3.7 & $771\pm28$ \\
\swift~/~00035088008 & XRT & 2011 Feb 08 & 4.6 & $558\pm24$ \\
\swift~/~00035088009 & XRT & 2011 Feb 13 & 3.1 & $466\pm22$ \\
\swift~/~00035088010 & XRT & 2011 Feb 15 & 1.7 & $202\pm14$ \\
\swift~/~00035088011 & XRT & 2011 Feb 20 & 3.8 & $689\pm26$ \\
\swift~/~00035088012 & XRT & 2011 Feb 27--28 & 4.9 & $941\pm31$ \\
\swift~/~00035088013 & XRT & 2011 Mar 27--28 & 6.9 & $1413\pm38$ \\
\swift~/~00035088014 & XRT & 2011 Apr 28 & 4.0 & $639\pm25$ \\
\swift~/~00035088015 & XRT & 2011 May 16 & 5.2 & $534\pm23$ \\
 & EPIC-pn & & 20.0 & $50600\pm227$\\
\emph{XMM}~/~0723570201 & EPIC-MOS\,1 & 2013 Sep 19 & 21.8 & $13085\pm116$\\
 & EPIC-MOS\,2 & & 21.7 & $5220\pm74$ \\
\hline
\end{tabular}
\begin{list}{}{}
\item[$^{a}$] Net source counts considering the source and background extraction regions described in the text. The counts are in the 0.3--10~keV band for the \swift\ XRT, in the 0.3--12~keV for the \xmm\ MOS cameras, and in the 1.5--12~keV for the \xmm\ pn camera.
\end{list}
\end{minipage}
\end{table*}

\subsection{\swift}

\src\ was observed by \swift\ 15 times between 2005 October and 2011 May, for a total net exposure of $\sim$80.8~ks (see Table\,\ref{logs}). The \swift\ XRT \citep{burrows05short} data were collected in imaging photon counting (PC) mode, with a CCD readout time of 2.507~s. The data were processed and reduced using standard software tools and procedures (\textsc{heasoft} v.~6.14~/~\textsc{caldb}~20130313). The source photons were selected within a 20-pixel radius (1 XRT pixel $\simeq$ 2.36 arcsec), while the background counts were accumulated from an annular region with radii of 50 and 80 pixels.

\subsection{\xmm}\label{xmmdata}
\xmm\ observed \src\ on 2013 September 19 for about 21~ks. The pn CCD camera \citep{struder01short} was set to operate in fast-timing mode (which achieves a time resolution of 0.03~ms by preserving only one-dimensional positional information), while the MOS\,1 and MOS\,2 CCD cameras \citep{turner01short} were run in small-window mode (time resolution: 0.3~s) and in full-frame mode (time resolution: 2.6~s), respectively. The data were processed and reduced using the \xmm\ \textsc{sas} v.~13.5 and the \textsc{ccf-rel-307} calibration release. The observation was affected by a few soft proton flares. The corresponding periods of high particle background were filtered out of the data using intensity filters and following the method described in \citet{deluca04}. 

To extract the pn source counts, we used a thin strip ($\sim$10 pixels) centred on the source and with the length covering the entire readout streak. The background spectrum was extracted from two symmetric 5-pixel-wide strips aside the source. The pn data suffered from the recurrence of instrumental noise bursts, which are characteristic of the fast-timing mode \citep{burwitz04}. These noise bursts are short ($\la$0.1~s) and have a very distinctive spectrum \citep{burwitz04}, with a peak around 0.20--0.25 keV, few counts above 0.3~keV, and essentially no counts above 1.5~keV. Given this, we preferred not to apply another good time-interval filter to the pn data to exclude these events, since this would have severely reduced the counting statistics and degraded the quality of the pn light curves. Instead, we limited the pn analysis to photons with energy above 0.5~keV for the timing analysis and above 1.5~keV for the spectral analysis.\footnote{In the pn light curves the peaks produced by the noise flares disappear above $\sim$0.3--0.4~keV, and we checked that residual spurious counts do not impact the timing analysis or alter the folded profiles. Indeed, this is not to be expected, since the noise bursts have typical duration ($\la$0.1~s) much shorter than the period of \src\ and are aperiodic. For the spectral analysis we adopted a more conservative selection, since residual spurious counts, being concentrated in a narrow energy range, might build up features in the spectra (especially when integrated over long exposures), and the flares cannot be properly subtracted from a nearby background region, since they are spatially inhomogeneous in the CCD chip.}

For the MOS\,1, we accumulated the source counts from a circular region with radius of 40~arcsec. Owing to the small field of the central CCD (CCD~\#1) in small-window mode, the background was estimated from a circular region in CCD~\#7, at a distance of about 6.7~arcmin from \src. The MOS\,2 data are affected by pileup. For this reason, we extracted the source photons from an annulus around \src, thus excluding the inner (piled-up) point spread function (PSF) core. We used 40~arcsec for the outer radius and $\sim$10~arcsec for the inner radius. The latter value was selected by fitting the PSF with a King function and using the \textsc{sas} task \textsc{epatplot}. The background was extracted from a circular region on the same chip as the source.

\section{Analysis and results}\label{analysis}

\begin{figure*}
\centering
\resizebox{\hsize}{!}{\includegraphics[angle=-90]{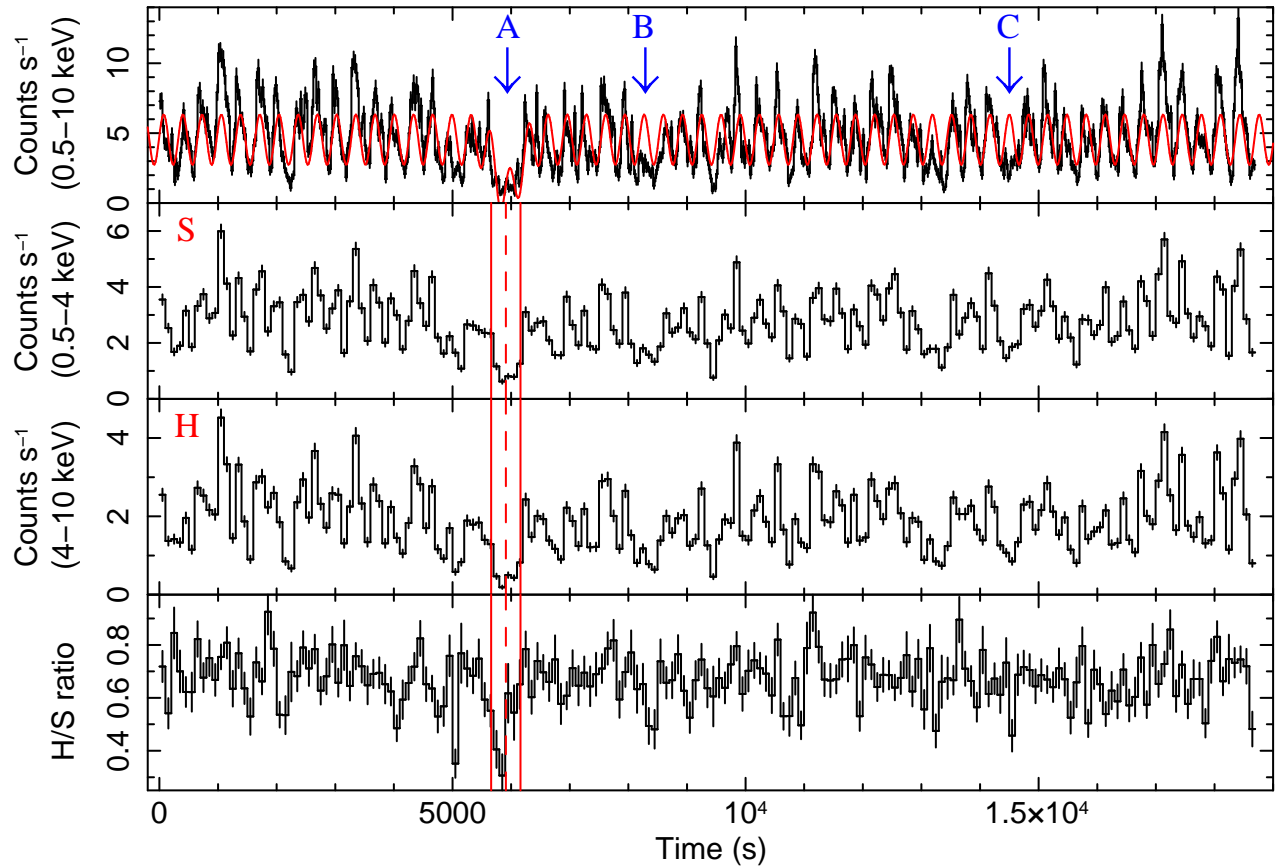}}
\caption{\label{lc} Top panel: \xmm\ EPIC 0.5--10 keV light curve of \src\ (bintime $=50$~s). We show only the time interval ($\sim$18.7~ks) with simultaneous pn, MOS\,1, and MOS\,2 coverage. The origin of time ($t_0$) is taken at 2013 September 19 03:41:12.8 (UTC). The red line is a sinusoidal fit to the average pulse profile and includes a Gaussian component to model the $\sim$400-s-long dip occurred during the observation (marked by arrow `A'). The arrows `B' and `C' indicate two other possible shorter dips. The background is negligible ($<$1\% of the total counts). Second panel: EPIC light curve in the soft (S) band 0.5--4~keV (bintime $=100$~s). Third panel: EPIC light curve in the hard (H) band 4--10~keV (bintime $=100$~s). Bottom panel: ratio between the hard (H) and soft (S) counts (errors were propagated by adding in quadrature). The vertical red lines in the last three panels indicate the time intervals used for the spectral analysis of the dip.}
\end{figure*}

\subsection{Light curves and timing analysis}\label{timingan}
As already mentioned, the periodic modulation in the X-ray emission of \src\ was detected during a run of the SATANASS\,@\,BAR project. Taking into account the whole sample of light curves analysed in the project ($\sim$4000), the peak at about 326.3~s in the Fourier power spectrum (which was computed from the first two \swift\ observations, collected in 2005 October 25--26) was significant at a 3.6$\sigma$ confidence level. Two harmonics were present in the power spectrum and the pulsed fraction of the signal was approximately 30\%. After the discovery of the periodicity, a \swift\ campaign was carried out to confirm the modulation and study the source. In Table\,\ref{timing}, we report the periods measured in the \swift\ data (note that few adjoining observations were combined) by means of a $Z^2_2$ analysis, together with the root-mean-square (rms) pulsed fractions.

The combined EPIC pn and MOS light curves from the 2013 \xmm\ observation are shown in Fig.\,\ref{lc} in the total (0.5--10~keV), soft (0.5--4~keV), and hard (0.5--4~keV) bands. The 328-s modulation is evident even without a folding analysis and the individual pulses have very variable amplitude. A dip, marked with the letter `A', is apparent in the light curve roughly 6~ks after the observation start [$(t-t_0)\approx6$~ks, where $t_0$ is fixed at 2013 September 19 03:41:12.8 (UTC)]. By a Gaussian fit, we estimated its full width at half-maximum (FWHM) length to be $402\pm16$~s (here and in the following, uncertainties are at 1$\sigma$ confidence level). 

Two `missing pulses' can be seen in the light curve around $(t-t_0)\approx 8.3$ and 14.5~ks (they are marked with the letters `B' and `C' in Fig.\,\ref{lc}). A similar feature is also present in the \swift\ observation 00035088002. While their short duration hinder a detailed analysis, they might be dips shorter than the `A' one, but of similar origin.

The hardness ratio of \src\ as function of the time (Fig.\,\ref{lc}, bottom panel) shows significant variations during the observation, as it is common in HMXBs (note that substantial hardness-ratio variations are observed also across the spin phase, see below). It is interesting to note that a sudden variation in the hardness ratio of \src\ occurred in coincidence of the dip `A'. On the other hand, other large variations, such as that observed around $(t-t_0)\approx 5$~ks, do not find correspondence in clear lacks of pulsations.

\begin{figure}
\centering
\resizebox{\hsize}{!}{\includegraphics[angle=0]{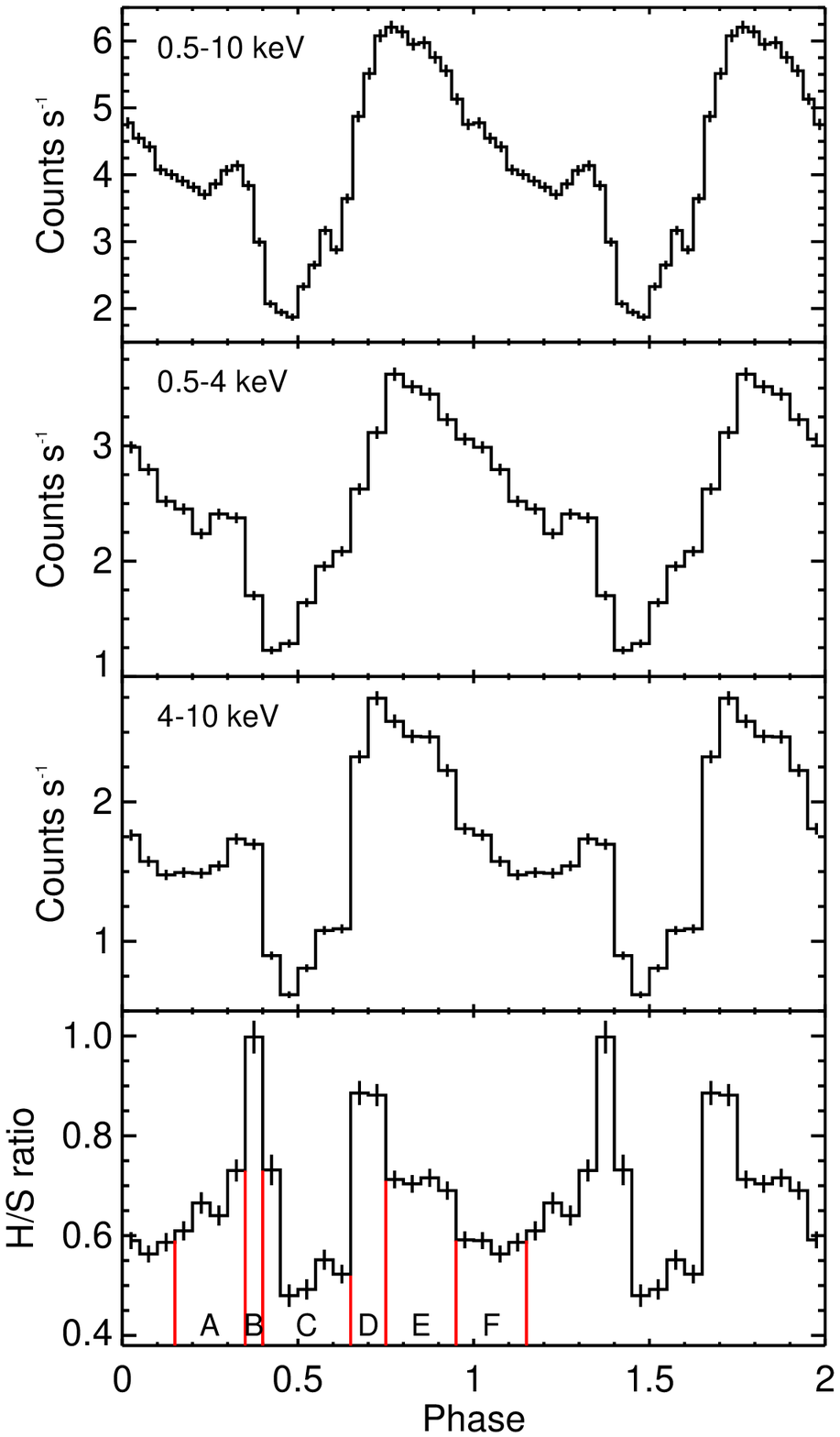}}
\caption{\label{xmmfold} The first three panels show the \xmm\ EPIC pulse profile of \src\ in the total, soft, and hard energy bands (as indicated in the panels). The total profile is displayed with 32 phase bins, the others with 20 phase bins. Bottom panel: ratio between the hard (H) and soft (S) profiles (errors were propagated by adding in quadrature). The red vertical lines indicate the phase intervals used for the phase-resolved spectroscopy.}
\end{figure}

In the EPIC data, we measured a period of $P=327.878\pm0.024$~s in the $Z^2_2$ periodogram. The folded pulse profile is shown in Fig.\,\ref{xmmfold} in three energy bands. It is evident that the profile changes as a function of energy. The rms pulsed fraction measured in the total band is $32.9\pm0.2$~\%, $29.2\pm0.3$~\% in the soft band, and $37.2\pm0.3$~\% in the hard band. For completeness, we computed the rms pulsed fractions also in finer energy bands: $28.8\pm0.8$~\% in 0.5--2~keV; $29.7\pm1.1$~\% in 2--3~keV; $34.5\pm1.2$~\% in 3--4~keV; $36.2\pm1.6$~\% in 4--5~keV; $37.1\pm1.7$~\% in 5--6~keV; $39.5\pm2.0$~\% in 6--7~keV; $38.7\pm2.7$~\% in 7--8~keV; $44.9\pm3.9$~\% in 8--9~keV; and $38.0\pm4.0$~\% in 9--10~keV.
\begin{table}
\centering \caption{Timing results.} \label{timing}
\begin{tabular}{@{}lcc}
\hline
Segment$^{a}$ & Period$^{b}$ & rms pulsed fraction$^{b}$ \\
 & (s) & (\%) \\
\hline
2005 Oct 26--27 & $326.276\pm0.016$ & $28.3\pm0.8$ \\
2010 Oct 22--27 & $328.178\pm0.005$ & $31.1\pm0.7$ \\
2011 Feb 03--28 & $328.1782\pm0.0009$ & $30.9\pm0.7$ \\
2011 Mar 27--28 & $328.16\pm0.05$ & $36.6\pm1.7$ \\
2011 Apr 28 & $327.65\pm0.54$ & $35.0\pm2.6$ \\
2011 May 16 & $327.96\pm0.28$ & $38.1\pm2.5$ \\
2013 Sep 19 & $327.878\pm0.024$ & $32.9\pm0.2$ \\
\hline
\end{tabular}
\begin{list}{}{}
\item[$^{a}$] See Table\,\ref{logs}. 
\item[$^{b}$] Periods were derived from a $Z^2_2$ test. Uncertainties were determined from Monte Carlo simulations.
\end{list}
\end{table}

\subsubsection{Other recent X-ray observations}

As mentioned before, \src\ was observed also by \cxo\ between 2007~September~30 and October~01 (obs.\,ID~7532, exposure time: 4.7~ks; \citealt{tomsick08}). The unabsorbed flux of \src\ was $\approx$$3\times10^{-11}$~\flux, causing significant pile-up in the \cxo\ ACIS-S instrument, which was operated in full-frame mode (frametime: 3.2~s). We searched the source data, considering a 2-arcsec radius region centred on \src, for the $\sim$327~s signal in the 0.3--8~keV energy band, but no significant pulsations were found (see also \citealt{nichelli11}). 
After discarding the piled-up events from the core of the \cxo\ PSF, a signal at $327.1\pm0.4$~s could be recovered (with significance of $\sim$4$\sigma$) by a folding analysis carried out in a $\pm$10~s interval centred at 327~s. Because of the poor counting statistics (about 200 source photons), the pulsed fraction cannot be constrained.

In 2003--2004, \src\ was serendipitously observed several times by \xte\ in a campaign devoted to the black hole X-ray transient XTE\,J1720--318 (see \citealt{brocksopp05}). In these observations, \src\ was always $\sim$$0\fdg5$ from the \xte\ pointing direction. A signal around 325~s is detected by a folding analysis (in the interval $327\pm10$~s) of the data of the longest observations with the \xte/PCA, which is a collimated instrument with a FWHM field of view of about 1$^\circ$. For instance, a period of $324.71\pm0.07$~s is observed on 2003~January~15 (obs.\,ID~70116-03-02-01, net exposure: 16.6~ks, $\sim$6$\sigma$ significance) and a period of $324.38\pm0.05$~s on 2003~February~20 (obs.\,ID~70123-01-03-07, net exposure: 9.8~ks, $\sim$5$\sigma$ significance). Considering the period variability observed in the \swift\ and \xmm\ data (Table\,\ref{timing}), these signals are likely due to \src, but the non-imaging nature of the \xte\ instruments preclude any definitive conclusions.

\subsection{Spectral analysis}

\subsubsection{Phase-averaged and phase-resolved spectral analysis}

\begin{figure}
\centering
\resizebox{\hsize}{!}{\includegraphics[angle=-90]{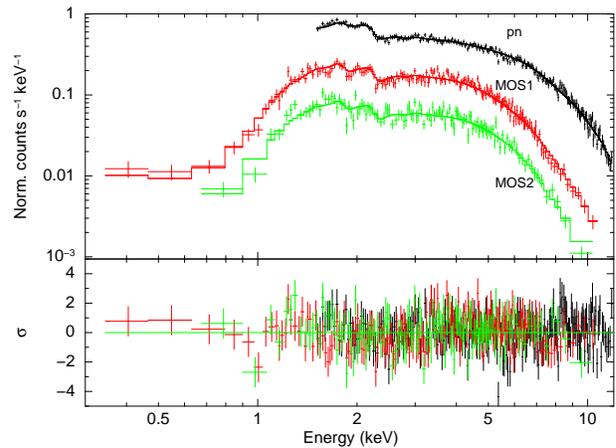}}
\caption{Fit of the \xmm\ spectra with the power law plus hot blackbody model. Bottom panel: residuals of the fit in units of standard deviations with error bars of size one.\label{xmm_spec}}
\end{figure}

For the phase-averaged spectral analysis, we first concentrate on the high counting statistics \xmm\ spectra. For the MOS spectra, we considered the energy range 0.3--12~keV, for the pn spectrum the band 1.5--12~keV (see Section\,\ref{xmmdata}). We fit to the EPIC spectra a set of models typically used for accreting pulsars (power law, power law plus blackbody, and cutoff power law, all modified for the photoelectric absorption). The summary of the spectral fits is given in Table\,\ref{specs}. All models indicate a rather hard spectrum, an observed 1--10~keV flux of $\sim$$2.75\times10^{-11}$~\flux, and an absorbing column of $(1$--$3)\times10^{22}$~cm$^{-2}$ (which is higher than the column density through the Galaxy along the line of sight, $\sim$$5\times10^{21}$~cm$^{-2}$ from \citealt{kalberla05}).
Apart from the power law, all the models tested fit reasonably well the \xmm\ data, but the blackbody plus power-law is the one that provides the best fit. Two variants of this model are consistent with the data: a relatively cool blackbody with $kT \sim0.15$~keV plus a power law with $\Gamma\sim1.2$, or a hotter blackbody with temperature corresponding to $\sim$1.2~keV and a harder power with photon index $\sim$0.8. In the following, we adopt the latter model (see Fig.\,\ref{xmm_spec}), which is better consistent with an HMXB interpretation of \src\ (see Section\,\ref{discussion}).

As shown in Fig.\,\ref{xmmfold}, where the pulse profiles in two energy bands (0.5--4 and 4--10~keV) and their ratio are displayed, the spectrum of \src\ is strongly variable with phase. In particular, the phase intervals marked with B and D in Fig.\,\ref{xmmfold} show the hardest X-ray emission, while the phase minimum (interval C) appears to be significantly softer. The pn and MOS spectra of the six phase intervals selected with the hardness ratio were extracted using the same procedure as for the average spectrum and fit with an absorbed power-law plus blackbody model (the one with $kT\simeq1.2$~keV in Table\,\ref{specs}). Forcing all the parameters to assume the same values for all the spectra, except for an overall normalization factor, a rather poor fit is obtained ($\chi^2_{\nu}=1.33$ for 2016 dof), with large residuals at low energies. An acceptable fit ($\chi^2_{\nu}=1.07$ for 2011 dof) is instead obtained by letting the absorption free to vary in each spectrum (even though this is not the only possible choice). The best-fitting \nh\ values (in units of $10^{22}$ cm$^{-2}$ and assuming the abundances from \citealt*{wilms00}) for each phase interval are the following: $1.61\pm0.09$, $3.4\pm0.2$, $1.12\pm0.08$, $2.89\pm0.12$, $1.66\pm0.09$, and $1.16\pm0.08$ for phase A, B, C, D, E, and F, respectively.

\subsubsection{Spectral analysis of the dip}

An indication of spectral variability of the source during the main dip (`A') clearly emerges from the hardness ratio curve shown in Fig.\,\ref{lc}. In particular, in the first part of the dip the source hardness reaches its minimum value. To better study this peculiar source state, we extracted the pn and MOS spectra from the full time interval of the dip (from about $t_0+5660$~s to  $t_0+6160$~s), its first and second half, and fit them with the best-fitting power law plus blackbody model (with $kT\simeq1.2$~keV, see Table\,\ref{specs}), rescaled by a normalization factor.\footnote{We note that a meaningful spectral analysis of the other possible dips (including the \swift\ one) is prevented by the very low number of counts.}

Despite the limited counting statistics (202 background-subtracted counts), the initial spectrum cannot be satisfactorily fit by this simple model ($\chi^2_{\nu}=3.07$ for 7 dof). Acceptable fits to the initial spectra are instead obtained with a variable absorption ($N_{\rm H}< 1.1\times10^{22}$ cm$^{-2}$ at 3$\sigma$; $\chi^2_{\nu}=1.67$ for 6 dof) or changing the parameters of either the power law ($\Gamma=2.4^{+1.7}_{-0.8}$; $\chi^2_{\nu}=0.78$ for 5 dof) or blackbody component ($kT=0.8\pm0.1$ keV; $\chi^2_{\nu}=0.55$ for 5 dof). The 1--10~keV unabsorbed fluxes are $6.0^{+0.5}_{-0.4}\times10^{-12}$~\flux\ for the fit with free \nh, $5.5^{+0.8}_{-0.5}\times10^{-12}$~\flux\ for the fit with free power-law component, and $5.3^{+1.2}_{-1.1}\times10^{-12}$~\flux\ in the case of free blackbody component.

In contrast, the spectrum of the second part of the dip (260 background-subtracted counts) and that extracted from the entire dip are adequately fit by the model of the phase-averaged spectrum with fixed parameters ($\chi^2_{\nu}=0.81$ for 9 dof and $\chi^2_{\nu}=1.36$ for 23 dof, respectively). The unabsorbed 1--10~keV flux is $(8.18\pm0.35)\times10^{-12}$~\flux\ when averaged along the whole dip length, and $(9.56\pm0.61)\times10^{-12}$~\flux\ in the second part of the dip.
We also tested the possibility of a complete suppression of the power-law component during the dip by fitting the corresponding spectrum with an absorbed blackbody model with \nh\ and temperature fixed at the best-fitting values of the power law plus (hot) blackbody fit to the average spectrum (Table\,\ref{specs}). This fit is not acceptable ($\chi^2_{\nu}=1.84$ for 23 dof) and shows systematic residuals at high energies, indicating the presence of a harder spectral component.

\begin{figure}
\centering
\resizebox{\hsize}{!}{\includegraphics[angle=-90]{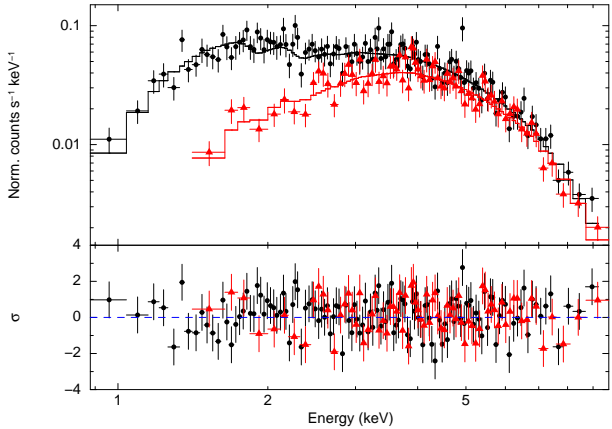}}
\caption{\label{xrtspec} Comparison of the \swift\ spectrum from observation 8003 with that obtained from the observations 8005, 8009, and 8010, which are very similar to each other, combined (red triangles). The solid lines show the best-fitting power-law models (same photon index, but different \nh\ and normalization). Bottom panel: residuals of the fit in units of standard deviations with error bars of size one.}
\end{figure}

The \swift\ spectra of the individual observations, collected over $\sim$6 years (see Table\,\ref{logs}), can be all described by a simple power law modified for the interstellar absorption.\footnote{Although models more complicated than a photoelectrically absorbed power law are necessary to properly fit the high-counting-statistics \xmm\ data, the addition of extra components did not improve the goodness of fit significantly for any of the \swift\ observations. Also considered that the emission of \src\ is variable, we opted to stick to this simple model to compare the spectral properties from \swift\ data set to data set.} They show, however, substantial variability in both flux and spectral shape  (see Fig.\,\ref{xrtspec}). In order to investigate whether changes in the absorbing column might account for part of the variations, we performed a simultaneous fit of the 15 \swift\ spectra with free \nh\ and normalizations, and the photon index tied between all the data sets. Indeed, they can be fit by a simple power law ($\chi^2_\nu=0.96$ for 763 dof) with photon index $\Gamma=1.03\pm0.03$ and \nh\ varying from $\sim$1.9 to $7\times10^{22}$~cm$^{-2}$ (see Fig.\,\ref{xrtlc}). We note that an almost as good fit is obtained when the hydrogen column value is tied between all observations and the other parameters are left free to vary ($\chi^2_\nu=1.04$ for 763 dof); in this case, the derived \nh\ is $(2.07\pm0.08)\times10^{22}$~cm$^{-2}$ and the photon index ranges from $\sim$0 to 1.2.
\begin{table*}
\begin{minipage}{17cm}
\centering \caption{\xmm\ spectral results. Errors are at a 1$\sigma$ confidence level for a single parameter of interest.} \label{specs}
\begin{tabular}{@{}lcccccccc}
\hline
Model$^a$ & \nh$^b$ & $\Gamma$ & $kT~/~E_\mathrm{c}$$^c$ & $E_\mathrm{f}$$^c$ & $R_{\mathrm{BB}}$$^d$ & Flux$^e$ & Unabs. flux$^e$ & $\chi^2_\nu$ (dof) \\
 & ($10^{22}$ cm$^{-2}$) &  & (keV) & (keV) & (km) & \multicolumn{2}{c}{($10^{-11}$ \flux)} & \\
\hline
\textsc{phabs\,(pl)} & $1.85\pm0.04$ & $1.08\pm0.01$ & -- & -- &  -- & $2.75\pm0.02$ & $3.23\pm0.02$ & 1.29 (509) \\
\textsc{phabs\,(cutoffpl)} & $1.51\pm0.06$ & $0.67\pm0.06$ & -- & $13\pm2$ &  -- & $2.74\pm0.02$ & $3.10\pm0.02$ & 1.19 (508) \\
\textsc{phabs\,(highecut\,*\,pl)} & $1.46\pm0.05$ & $0.80\pm0.04$ & $3.5\pm0.2$ & $16^{+3}_{-1}$ &  -- & $2.74\pm0.02$ & $3.09\pm0.02$ & 1.14 (507) \\
\textsc{phabs\,(bb\,+\,pl)} & $1.31\pm0.07$ & $0.81\pm0.06$ & $1.18\pm0.05$& -- & $0.27\pm0.03$ & $2.74\pm0.02$ & $3.04\pm0.02$ & 1.12 (507) \\
\textsc{phabs\,(bb\,+\,pl)} & $2.84\pm0.13$ & $1.22\pm0.02$ & $0.15\pm0.01$& -- &  $88^{+17}_{-14}$ & $2.74\pm0.02$ & $3.95\pm0.10$ & 1.12 (507) \\
\hline
\end{tabular}
\begin{list}{}{}
\item[$^{a}$] \textsc{xspec} models; \textsc{bb = bbodyrad, pl = powerlaw}.
\item[$^{b}$]  We used the abundances of \citet*{wilms00} and the photoelectric absorption cross-sections from \citet{balucinska92}.
\item[$^{c}$] $E_\mathrm{c}$: cutoff energy; $E_\mathrm{f}$: $e$-folding energy.
\item[$^{d}$] The blackbody radius is calculated at infinity and for an arbitrary distance of 5~kpc.
\item[$^{e}$] In the 1--10 keV energy band.
\end{list}
\end{minipage}
\end{table*}

\section{Discussion}\label{discussion}

We reported on an in-depth characterization at X-rays of the source \src, both from the temporal and the spectral point of view. Pulsations, originally discovered by our group in \swift\ data and  reported by \citet{nichelli11}, have been recovered in an archival \cxo\ observation, and clearly observed in the new \xmm/EPIC data analysed in detail here, with a periodicity of $327.878\pm0.024$~s. Such a long pulse period is typical of a neutron star in an HMXB. In fact, an early-type companion is also indicated by optical studies \citep{masetti06short} and by the precise \cxo\ position \citep{tomsick08}.
Both a Be (main sequence or giant) star and a blue supergiant are viable possibilities for the companion star of \src. Unfortunately, for \citet{masetti06short} it was not possible to derive significant information about the optical counterpart (spectral type, absorption and source distance), since a reliable photometry was missing. From the Corbet diagram of X-ray-pulsar spin period versus orbital period \citep{corbet86}, the observed pulse period suggests an orbital period of either $\sim$3--30~d or $\sim$100--200~d for an HMXB hosting either a supergiant donor or a Be, respectively. Strong pulse to pulse variations (usually produced by fluctuations in the wind accreted by the neutron star) and a pulse profile energy dependence are evident in \src, as it is often observed in accreting pulsars with massive donors.

The hard \xmm\ spectrum (power-law photon index in the range 0.8--1.2) is in line with an HMXB origin for the X-ray emission. The presence of a soft component is indicated by the X-ray spectroscopy. The spectrum is well fit by a blackbody emission together with a hard power law. While both a colder ($kT\sim0.15$~keV) and a hotter ($kT\sim1.18$~keV) blackbody are valid deconvolution of the spectrum, we favour the hot solution, which is consistent in temperature and size ($R_{\mathrm{BB}}\approx0.3\,d_5$~km, where $d_5$ is the distance in units of 5~kpc) with an HMXB interpretation of the hot blackbody as produced in the polar cap region of the neutron star accretion column \citep{becker07}.  

\begin{figure}
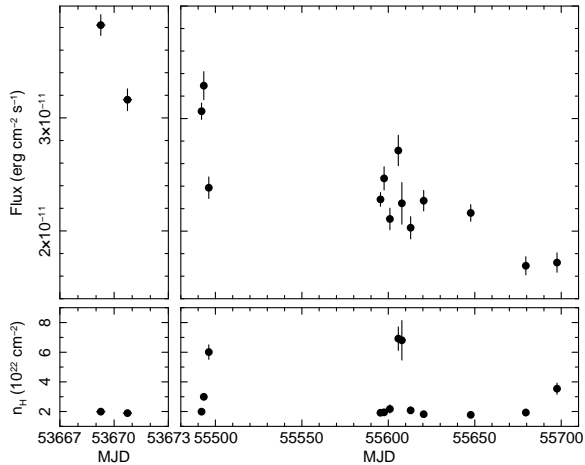

\centering
\resizebox{\hsize}{!}{\includegraphics[angle=-90]{part1.ps}\hspace{-2.5cm}\includegraphics[angle=-90]{part2.ps}}
\caption{\label{xrtlc} Top panels: \swift\ long-term light curve (the flux is in the 1--10 keV energy range and not corrected for the absorption). Note the time gap and the different time-scales. Bottom panels: corresponding absorption column as inferred from the spectral analysis.}
\end{figure}

The strong variability of the spin-phase-selected spectra (Fig.\,\ref{xmmfold}, bottom panel) can be accounted for by a variable absorbing column density along the spin cycle, probably mapping a different density of the local matter illuminated by the X-ray beam pattern. Significant variations (up to a factor of 4) in the absorbing column density are also present in the long-term X-ray light curve (Fig.\,\ref{xrtlc}). Variations in the local absorption are usual in HMXBs because the compact object is constantly embedded in an intense structured and clumpy wind. Moreover, in a few cases, large-scale structures have been observed in HMXBs (gas streams) and interpreted as the result of the disruption of the wind by the neutron star passage, which produce density perturbations that modulate the observed \nh\ along the orbital cycle \citep{blondin90,manousakis11}.

We observed a remarkable feature in the EPIC light curve, a dip or a so-called off-state, characterized by a reduction in the source intensity down to $\sim$20--30\% of its normal level, and lasting for about one neutron star spin cycle. Similar features have been seen so far only in a few HMXB pulsars with supergiant companions and are potentially important to derive information on the accretion regime and the neutron star properties: \mbox{Vela\,X--1} (orbital period $P_{\mathrm{orb}}\simeq9$~d and spin period $P_{\mathrm{spin}}\simeq283$~s; \citealt{inoue84,kreykenbohm99,kreykenbohm08short}), 4U\,1907+09 ($P_{\mathrm{orb}}\simeq8.4$~d and $P_{\mathrm{spin}}\simeq437.5$~s; \citealt{dsdk12}), GX\,301--2 ($P_{\mathrm{orb}}\simeq41.5$~d and $P_{\mathrm{spin}}\simeq686$~s; \citealt*{gkb11}), and in the Supergiant Fast X-ray Transients IGR\,J16418--4532 ($P_{\mathrm{orb}}\simeq3.7$~d and $P_{\mathrm{spin}}\simeq1212$~s; \citealt{drave13}), and  IGR\,J17544--2619 ($P_{\mathrm{orb}}\simeq4.9$~d and candidate $P_{\mathrm{spin}}\simeq71.5$~s; \citealt{drave14}). During these off-states, X-ray pulsations sometimes appear to be suppressed  \citep{kreykenbohm99}, sometimes are still detected \citep*{doroshenko11}, and the X-ray spectrum softens. This implies that these dips are not caused by the obscuration by a dense wind clump passing our line of sight to the pulsar, but are due to a real flux drop. \citet{gkb11} suggested that softer X-rays during these dips are likely due to the suppression of the harder X-ray emission produced in the accretion column, leaving visible the soft X-rays coming from the underlying thermal mound at the bottom of the accretion column \citep{becker07}. The reason for a suppression (or cessation) of the hard X-rays produced in the accretion column is unclear.
Aside from temporarily enhanced absorption along the line of sight, three main explanations for off-states have been proposed: a temporary transition to a centrifugal inhibition of the accretion (see e.g. \citealt{kreykenbohm08short} for Vela\,X--1), a Kelvin--Helmholtz (KH) instability \citep{dsdk12}, or an accretion regime change from a Compton cooling regime (producing a higher luminosity) to a radiative cooling regime (lower luminosity) in the equatorial plane of the neutron star magnetosphere, due to a switch from fan beam to pencil beam emission pattern \citep{shakura13}.

Also in \src, the significantly softer spectrum during the off-state leads us to exclude the possibility of an obscuration of the central source by a wind structure (or clump), which would cause instead a spectral hardening. The small duration of the dip (a little more than a spin cycle) does not allow us to establish if the remaining fainter emission is pulsating, so a centrifugal barrier temporarily halting accretion, could be a possibility, although the short duration of the off-state makes it unlikely \citep{gkb11}. In \src, the spectrum during the dip cannot be completely accounted for {\em only} by an unvarying blackbody emission (this is ruled out by the marked spectral variability), and significant (power-law-like) hard X-ray emission is present, suggesting that in our case accretion is not suppressed in contrast, again, to what is expected with the onset of a propeller regime.

A transition to accretion via KH instability is discussed by \citet{dsdk12} as a possible explanation of dips in the transient source 4U\,1907+09, following the theory of wind accretion in HMXBs surveyed by \citet*{bozzo08}. In the framework of the different accretion regimes discussed by \citet{bozzo08}, the magnetospheric boundary at the Alfv\'en radius, $R_{\mathrm A}$, can be KH unstable in two regimes: in the sub-Keplerian magnetic inhibition regime (where $R_{\mathrm{acc}} < R_{\mathrm{A}} < R_{\mathrm{cor}}$, where $R_{\mathrm{acc}}$ is the accretion radius, $R_{\mathrm{cor}}$ is the corotation radius, where the neutron star angular velocity is equal to the Keplerian angular velocity), and in the subsonic propeller regime (where $R_{\mathrm{A}} < R_{\mathrm{acc}} < R_{\mathrm{cor}}$). The spin period of $\sim$328~s implies a corotation radius, $R_{\mathrm{cor}}$, of $8\times 10^{9}$~cm. Thus, in both regimes, $R_{\mathrm{cor}}$ represents an upper limit to the value of the other two important radii involved. The relation $R_{\mathrm{acc}} < R_{\mathrm{cor}}$ ($R_{\mathrm{acc}}=2GM/v_{\mathrm{rel}}^2$, where $M$ is the neutron star mass, and $v_{\mathrm{rel}}$ is the relative velocity of the neutron star and the companion wind) translates into a lower limit for $v_{\mathrm{rel}}$ of $\sim$2200~km~s$^{-1}$, which is, on average, quite high in comparison with usual HMXBs, but it cannot be excluded, given the large variability expected in inhomogeneous winds of massive stars \citep*{oskinova12}.

Adopting equation~21 in \citet{bozzo08} for the X-ray luminosity $L_{\rm KH}$ produced by matter entering the magnetosphere through KH instability in the subsonic propeller regime, all the parameters involved in this formula are basically unknown for \src, except for the neutron star spin period. However, we can derive an upper limit to $L_{\rm KH}$ by assuming the following: $R_{\mathrm{acc}} = R_{\mathrm{A}} = R_{\mathrm{cor}}$, an orbital period of 5~d (which is reasonable for an accreting pulsar with a supergiant companion and the observed spin period; \citealt{corbet86}) and a total mass for the binary system of 30~M$_{\odot}$ (implying $a_{\mathrm{10d}}=0.63$ in equation~21 in \citealt{bozzo08}), a relative velocity of 2200~km~s$^{-1}$ ($v_8=2.2$ in equation~21 in \citealt{bozzo08}), a wind mass-loss rate of $10^{-6}$~M$_{\odot}$~yr$^{-1}$, a density ratio $\rho_{\mathrm{i}}/\rho_{\mathrm{e}} = 1$ [where $\rho_{\mathrm{i}}$ and $\rho_{\mathrm{e}}$ are the internal (below $R_{\mathrm{A}}$) and the external densities (above $R_{\mathrm{A}}$)], and $\eta_{\rm KH}\sim0.1$ for the  efficiency factor.
This translates into $L_{\rm KH} < 1.5\times10^{35}$~\lum, which can easily be accounted for if one assumes a reasonable distance to the source (see below). Note that the upper limit to the Alfv\'en radius, $R_{\mathrm{A}} <  R_{\mathrm{cor}}$, results in an upper limit to the dipolar magnetic field of  $B<1.2\times10^{14}$~G ($\mu<6\times 10^{31}$~G~cm$^{3}$; equation~19 in \citealt{bozzo08}). Again, we caution that all these estimates rely on particular (though reasonable) assumptions on all other parameters involved in the system geometry and wind properties. 
Even larger uncertainties are involved in the estimate of the KH instability mass accretion rate across the magnetosphere in the sub-Keplerian magnetic inhibition regime, where also the shear velocity is involved. Assuming the same values as above for the parameters in equation~10 in \citet{bozzo08}, we derive $L_{\rm KH}\sim10^{34}$~\lum.

A third possible explanation for off-states involves the application of the \citet{shakura12} model of subsonic quasi-spherical accretion on to moderately low luminosity ($<$$4\times10^{36}$~\lum) and slowly rotating neutron stars. Here, off-states are explained by transitions to an ineffective accretion regime, rather than with a cessation of the accretion. In this scenario, an off-state is the signature of a transition from a regime where the cooling of the gravitationally captured plasma entering the neutron star magnetosphere is dominated by Compton processes to a less efficient radiative cooling regime \citep*{shakura13}. 
This transition  might be triggered by a change in the X-ray beam pattern, from a fan beam to a pencil beam, produced by a reduced optical depth in the accretion flow. Since matter enters the neutron star magnetosphere more easily from the equatorial region, a high lateral X-ray luminosity by a fan beam directly illuminating the equator, allows a more efficient Compton cooling, facilitating accretion and thus resulting in a higher luminosity. So, a transition from a fan to a pencil X-ray beam pattern increases the Compton cooling time, triggering a lower luminosity state (an off-state). The transition back to the normal X-ray luminosity can be triggered by an enhanced density above the magnetosphere, increasing the accretion rate.
The transition to an off-state occurs at an X-ray luminosity of about $3\times 10^{35} \mu_{30}^{-3/10}$~\lum, where $\mu_{30}=\mu/[10^{30}$~G~cm$^3]$ is the neutron star dipole magnetic moment. The luminosity level in the off-state is about $10^{35} \mu_{30}^{7/33}$~\lum\ in the radiation cooling regime \citep{shakura13}. 

The average unabsorbed 1--10~keV flux of \src\ during the \xmm\ observation varies, depending on the adopted model (see Table\,\ref{specs}), from $\sim$$3\times10^{-11}$ to $\sim$$4\times10^{-11}$~\flux. Nothing is known about the distance of \src, which could be anywhere towards the edge of the Galaxy. At 5~kpc, the former flux translates into a luminosity of $9\times10^{34}d^2_5$~\lum, the latter into $1.2\times10^{35}d^2_5$~\lum\ (where $d_5$ is the distance in units of 5~kpc). Above $\approx$$4\times10^{36}$~\lum, the quasi-spherical shell cannot exist and supersonic (Bondi) accretion is more likely, but this would require a distance $d>28$~kpc. This implies that, for reasonable source distances in the range 5--10 kpc, the X-ray emission easily matches the X-ray luminosity needed to trigger the switch to an off-state, assuming a standard neutron star magnetic field of $\sim$$10^{12}$~G. 

To conclude, the observed spectral softening during the dip (which is common in similar off-states observed in other HMXBs) is more naturally explained by a transition to an ineffective accretion regime, instead of an obscuring dense clump in front of the X-ray source (which would have implied an hardening). However, given the unknown distance, the different possibilities for the transition to an inefficient accretion regime discussed above cannot be clearly disentangled.

\section*{Acknowledgements} 
Based on observations obtained with \swift\ and \xmm. \swift\ is a NASA mission with participation of the Italian Space Agency and the UK Space Agency. \xmm\ is an ESA science mission with instruments and contributions directly funded by ESA Member States and NASA. This research has also made use of data collected with the \cxo\ and \xte\ satellites, which were obtained from the NASA's HEASARC archive.

\bibliographystyle{mn2e}
\bibliography{biblio}

\bsp

\label{lastpage}

\end{document}